\documentclass{elsart3}
\journal{New Astronomy}

\usepackage{natbib}
\usepackage{graphicx}
\usepackage{amssymb}

\def\simlt{\lower.5ex\hbox{$\; \buildrel < \over \sim \;$}}
\def\simgt{\lower.5ex\hbox{$\; \buildrel > \over \sim \;$}}
\def\hst{{\em HST}}
\def\hnot{\ifmmode H_0 \else H$_0$ \fi}

\def\msun{\ifmmode {\rm M_\odot} \else M$_\odot$\fi}
\def\lsun{\ifmmode {\rm L_\odot} \else L$_\odot$\fi}

\def\deg{\ifmmode ^{\circ}
         \else $^{\circ}$\fi}
\def\pdeg{\ifmmode
           $\setbox0=\hbox{$^{\circ}$}\rlap{\hskip.11\wd0 .}$^{\circ}
     \else \setbox0=\hbox{$^{\circ}$}\rlap{\hskip.11\wd0 .}$^{\circ}$\fi}
\def\msunyr{\ifmmode {\rm M_\odot~yr^{-1}}\else${\rm M_\odot~yr^{-1}}$\fi}
\def\lam{\ifmmode {\lambda} \else {$\lambda$} \fi}
\def\lamLlam{\ifmmode \lambda L_{\lambda}(5100) \else {$\lambda L_{\lambda}(5100)$} \fi}
\def\nulnu{\ifmmode \nu L_{\nu}(5100) \else {$\nu L_{\nu}(5100)$} \fi}

\def\mdoto{\ifmmode {\dot{M}_0} \else  $\dot{M}_0$ \fi}
\def\teff{\ifmmode {T_{eff}} \else $T_{eff}$ \fi}
\def\ilam{\ifmmode {I_\lambda} \else  $I_\lambda$ \fi}
\def\inu{\ifmmode {I_\nu} \else  $I_\nu$ \fi}
\def\fnu{\ifmmode {F_\nu} \else  $F_\nu$ \fi}
\def\yr{\ifmmode {\rm yr} \else  yr \fi}
\def\cm{\ifmmode {\rm cm} \else  cm \fi}
\def\cmmitwo{\ifmmode \rm cm^{-2} \else $\rm cm^{-2}$\fi}
\def\cmmithree{\ifmmode \rm cm^{-3} \else $\rm cm^{-3}$\fi}
\def\cmps{\ifmmode \rm cm~s^{-1}\else $\rm cm~s^{-1}$\fi}
\def\cmpsps{\ifmmode \rm cm~s^{-2}\else $\rm cm~s^{-2}$\fi}
\def\kmps{\ifmmode \rm km~s^{-1}\else $\rm km~s^{-1}$\fi}
\def\kmpspmpc{\ifmmode \rm km~s^{-1}~Mpc^{-1} \else
    $\rm km~s^{-1}~Mpc^{-1}$\fi}
\def\ergps{\ifmmode \rm erg~s^{-1} \else $\rm erg~s^{-1}$ \fi}
\def\ergpspcm{\ifmmode \rm erg~s^{-1}~cm^{-2} \else $\rm erg~s^{-1}~cm^{-2}$ \fi}
\def\ergpspcmphz{\ifmmode \rm erg~s^{-1}~cm^{-2}~Hz^{-1} \else $\rm
erg~s^{-1}~cm^{-2}~Hz^{-1}$ \fi}
\def\ergpspcmpa{\ifmmode \rm erg~s^{-1}~cm^{-2}~\AA^{-1} \else $\rm
erg~s^{-1}~cm^{-2}~\AA^{-1}$ \fi}
\def\ergpsphz{\ifmmode \rm erg s^{-1} Hz^{-1} \else
   $\rm erg s^{-1} Hz^{-1}$ \fi}
\def\mbh{\ifmmode M_{\mathrm{BH}} \else $M_{\mathrm{BH}}$\fi}
\def\msigma{\ifmmode M_{\sigma} \else $M_{\sigma}$\fi}
\def\mbulge{\ifmmode M_{\mathrm{bulge}} \else $M_{\mathrm{bulge}}$\fi}
\def\mgal{\ifmmode M_{\mathrm{gal}} \else $M_{\mathrm{gal}}$\fi}
\def\mgalstar{\ifmmode M^*_{\mathrm{gal}} \else $M^*_{\mathrm{gal}}$\fi}

\def\mbhsigstar{\ifmmode M_{\mathrm{BH}}~-~\sigma_* \else
  $M_{\mathrm{BH}}~-~\sigma_*$ \fi} 
\def\deltalogmbh{\ifmmode \mbox{\Delta~{\mathrm{log}}~M_{\mathrm{BH}}} \else \mbox{$\Delta$~log~$M_{\mathrm{BH}}}$\fi}

\def\sigstar{\ifmmode \sigma_* \else $\sigma_*$\fi}
\def\sigthree{\ifmmode \sigma_{\mathrm{[O~III]}} \else $\sigma_{\mathrm{[O~III]}}$\fi}
\def\sigtwo{\ifmmode \sigma_{\mathrm{[O~II]}} \else $\sigma_{\mathrm{[O~II]}}$\fi}
\def\wthree{\ifmmode {\rm FWHM({[O~III]})} \else $FWHM({[O~III]})$ \fi}
\def\wtwo{\ifmmode {\rm FWHM({[O~II]})} \else $FWHM({[O~II]})$ \fi}
\def\mthree{\ifmmode M_{\mathrm [O~III]} \else $M_{\mathrm [O~III]}$ \fi}
\def\mtwo{\ifmmode M_{\mathrm [O II]} \else $M_{\mathrm [O II]}$ \fi}
\def\log{\mathrm{log}\,}
\def\led{\ifmmode L_{\mathrm{Ed}} \else $L_{\mathrm{Ed}}$\fi}
\def\lbol{\ifmmode L_{\mathrm{bol}} \else $L_{\mathrm{bol}}$\fi}

\def\msun{\ifmmode {\rm M_\odot} \else M$_\odot$\fi}
\def\lsun{\ifmmode {\rm L_\odot} \else L$_\odot$\fi}
\def\mbh{\ifmmode M_{\mathrm{BH}} \else $M_{\mathrm{BH}}$\fi}
\def\msig{\ifmmode M_{\sigma} \else $M_{\sigma}$\fi}
\def\sigstar{\ifmmode \sigma_* \else $\sigma_*$\fi}
\def\mbhsigstar{\ifmmode M_{\mathrm{BH}}-\sigma_* \else $M_{\mathrm{BH}}-\sigma_*$ \fi}
\def\deltalogmbh{\ifmmode \Delta~{\mathrm{log}}~M_{\mathrm{BH}} \else $\Delta$~log~$M_{\mathrm{BH}}$\fi}
\def\sigthree{\ifmmode \sigma_{\mathrm{[O~III]}} \else $\sigma_{\mathrm{[O~III]}}$\fi}
\def\sigtwo{\ifmmode \sigma_{\mathrm{[O~II]}} \else $\sigma_{\mathrm{[O~II]}}$\fi}
\def\mthree{\ifmmode M_{\mathrm [O~III]} \else $M_{\mathrm [O~III]}$ \fi}
\def\mtwo{\ifmmode M_{\mathrm [O II]} \else $M_{\mathrm [O II]}$ \fi}
\def\hbeta{\ifmmode {\rm H}\beta \else H$\beta$\fi}
\newcommand{\civ}{{\sc [C~iv]}}
\newcommand{\mgii}{Mg~{\sc ii}}
\newcommand{\oiii}{{\sc [O~iii]}}
\newcommand{\oii}{{\sc [O~ii]}}
\newcommand{\feii}{{[Fe~{\sc ii]}}}
\def\kmps{\ifmmode \rm km~s^{-1}\else $\rm km~s^{-1}$\fi}

\begin{document}

\begin{frontmatter}



\title{Evolution of the Black Hole -- Bulge Relationship in QSOs}


\author{ G.~A. Shields$^1$, S. Salviander$^1$, E.~W. Bonning$^2$}

\address{1. Department of Astronomy, University of Texas, Austin, Texas, USA;
shields@astro.as.utexas.edu\\
2. Laboratoire de l'Univers et de ses Th\'{e}ories, Observatoire de Paris, F-92195 Meudon Cedex, France}

\runtitle{Black Hole -- Bulge Relationship}
\runauthor{G. Shields}

\begin{abstract}

QSOs allow study of the evolution of the relationship between black
holes in galactic nuclei and their host galaxies.  The black hole mass
\mbh\ can be derived from the widths of the broad emission lines, and
the stellar velocity dispersion \sigstar\ of the host galaxy can be inferred from
the narrow emission lines.  Results based on \oiii\ and \oii\  line widths
indicate that the \mbhsigstar\ relationship,  at redshifts  up to $z \approx 2$, is 
consistent with  no evolution or an increase of up to $\sim0.5$~dex in \mbh\ 
at fixed \sigstar.  CO line widths offer an estimate of \sigstar\ for luminous
QSOs at high redshifts.  The available objects from $z \approx 4$ to 6 have
very massive black holes, $\mbh \sim 10^{9.5}~\msun$, but their CO
line widths suggest much smaller host galaxies than would be expected
by the local \mbhsigstar\ relationship. The most massive black holes must
continue to reside in comparatively modest galaxies today, because
their number density inferred from QSO statistics exceeds the
present-day abundance of proportionally massive galaxies. 

\end{abstract}


\end{frontmatter}

\section{Introduction}
\label{intro}

The study of black hole demographics has added a new dimension to 
research involving active galactic nuclei (AGN).  This is rooted in
two developments of recent years.  The first is the availability of
measurements of supermassive black holes in nearby galaxies,
involving observations of stellar and gaseous motions with \hst\ along
with other techniques (reviews by Kormendy \& Gebhardt 2001; Ferrarese
\& Ford 2004; Combes 2005).   This has led to
the realization that \mbh\ is closely correlated with the luminosity
and especially the velocity dispersion of the bulge component of the
host galaxy (Gebhardt et al. 2000; Ferrarese \& Merritt 2000).   

The local \mbhsigstar\ relationship is given by
Tremaine et al. (2002) as 
\begin{equation}
\label{e:tremaine}
M_{\mathrm{BH}} = (10^{8.13} M_{\odot})(\sigma_*/200~\mathrm{km~s}^{-1})^{4.02}.
\end{equation}
The rms dispersion of only $\sim0.3$ in $\mathrm{log}\,\mbh$ in this
relationship suggests a fundamental connection between the evolution
of supermassive black holes and their host galaxies.  The formation
and evolution of supermassive black holes has become a focus of
theoretical study (review by Haiman \& Quataert 2004; Croton
2005, and references therein). 

The second development is the ability to estimate \mbh\ in
AGN based on evidence that the broad emission lines come from gas
orbiting the black hole at a radius that can be estimated from the
continuum luminosity (Kaspi et al. 2005, and references therein).
Central black hole masses in AGN can easily be
estimated in this fashion, allowing demographic studies and providing
insight into AGN physics. 
Results below assume a cosmology with $H_0 = 70~\kmpspmpc$, $\Omega_M = 0.3$,
and $\Omega_\Lambda = 0.7$.

\section{Black Holes in QSOs}
\label{bhqso}

 Direct measurements of nuclear black holes, based on spatially
 resolved measurements of the velocities of stars and gas within the
 gravitational sphere of influence of the hole,  are limited to nearby
 objects with negligible look-back times.   QSOs afford an
 opportunity to study the \mbhsigstar\ relationship as a function of
 cosmic time.  However, this requires a
 measurement of the luminosity or velocity dispersion of the host
 galaxy  in addition to the black hole mass.  The host galaxy
 luminosity and the stellar velocity 
 dispersion are difficult to measure directly at high redshift, given
 the glare of the active nucleus.  An alternative approach is to
 estimate \sigstar\ through the use of some surrogate involving
 emission lines of gas orbiting in the host galaxy. 
 
Nelson \& Whittle (1996) found that, on average, the width
of \oiii\ tracks the stellar velocity dispersion, $\sigthree \equiv
\wthree/2.35~\approx~\sigstar$.  This is supported by the agreement of
\mbh\ and \sigthree\ in nearby AGN with the local \mbhsigstar\
relationship (Nelson 2000).  Further support comes from the study of
Bonning et al. (2005), who found overall agreement between \sigthree\
in low redshift QSOs and the value of \sigstar\ implied by the
measured host galaxy luminosity and the Faber-Jackson relation.  The
rms scatter is $\sim0.13$  in $\log\,\sigthree$ 
at fixed host luminosity.  This suggests that \sigthree\ may be a
useful proxy for \sigstar\ in statistical studies, although not for
individual objects. 

\begin{figure}[!t]
 \includegraphics[width=\columnwidth]{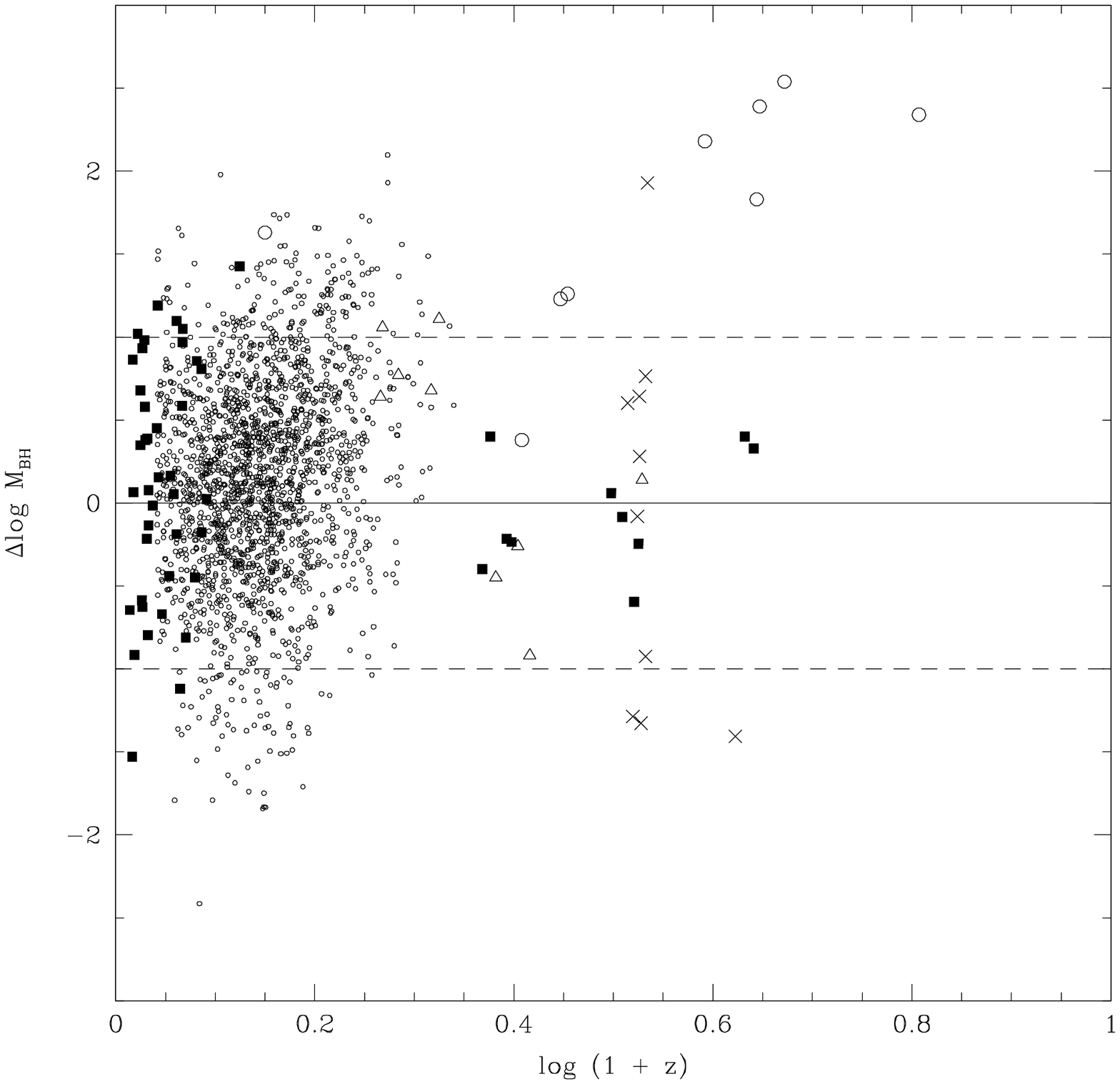}
   \caption{
Deviations from the local \mbhsigstar\ relationship, \deltalogmbh, as a function of redshift.
Filled squares: \hbeta~-~\oiii\  results from Shields et al. (2003);
small circles:  \hbeta~-~\oiii\ and
\mgii~-~\oii\  results from Salviander et al. (2006); 
open triangles: \oiii\ measured from spectra by Sulentic et al. (2004, see text and Table 1);  
crosses:  \hbeta~-~\oiii\  results from Netzer et al. (2004) and
Shemmer et al. (2004); 
large open circles: high redshift CO QSOs from Shields et al. (2006), who suggest a recalibration downward by $\sim0.5$ from the values shown. }
   \label{f:dmbh}
\end{figure}

Shields et al. (2003) carried out such a program using the
narrow \oiii\  $\lambda5007$ emission line of QSOs.  They collected
measurements of the widths of the broad \hbeta\ line and narrow \oiii\
line in QSOs ranging in redshift from $z < 0.1$ to $z = 3.3$.  Black
hole masses were derived from the expression 
\begin{equation}
\label{e:mbh}
\mbh = (10^{7.69}~M_{\odot})v_{3000}^2 L_{44}^{0.5},
\end{equation}
where $v_{3000} \equiv$ FWHM(\hbeta)$/3000$ km s$^{-1}$ and $L_{44}
\equiv \nu L_{\nu} /(10^{44}$ erg s$^{-1})$, the continuum luminosity
at 5100~\AA, based on Kaspi et al. (2000).
They expressed their results in terms of the deviation of an object's
actual \mbh\ from the value 
\msigma\ expected on the basis of Equation \ref{e:tremaine} using
\sigthree.  We follow their use of
$\deltalogmbh \equiv \log\,\mbh~-~\log\,\msigma$.  Figure \ref{f:dmbh} shows the results for  
\deltalogmbh\ as a function of redshift.  There is considerable
scatter, but overall the objects show little systematic offset from
Equation \ref{e:tremaine} as a function of redshift.  Shields et
al. concluded that \mbh\ at $z\sim2$ differs by less than 0.5~dex from
the present day value expected for a given \sigstar.  However, the
high redshift objects have 
larger masses ($\mbh \approx 10^{9.6}~\msun$) than the low redshift objects
($\sim10^{8.3}~\msun$), so the evolutionary comparison involves some disparity in mass.

Salviander et al. (2006; see also these proceedings) used these methods to assess the evolution of the \mbhsigstar\ relationship in the QSOs in Data Release 3 (DR3) of the 
Sloan Digital Sky Survey\footnote{The SDSS Web site is
  http://www.sdss.org/.} (SDSS; Abazajian et al. 2005).   They used
\hbeta\ and \oiii\ for redshifts up 
to 0.8, and the broad \mgii~$\lambda2800$ and 
narrow \oii~$\lambda3727$ emission lines to reach redshifts up to 1.2.
Their results, 
included in Figure \ref{f:dmbh}, nominally show a rise of $\sim0.5$ in
\deltalogmbh\ 
from $z \sim 0.3$ to $z \sim 1.0$.  However, they show that two
sources of bias contribute to this apparent evolution. (1) The scatter
in the local \mbhsigstar\ relationship causes some galaxies to harbor
exceptionally large black holes.  If these are fueled in proportion to
their mass, their high luminosities cause them to be over-represented
in the QSO sample. 
(2)  There is a tendency in data of marginal signal/noise quality to
fail to detect wider lines, over-representing narrower \oiii\ lines in
the sample.  Subtracting these biases, which both act to give an
increase in \deltalogmbh\ with redshift, 
Salviander et al. find a residual evolution of \mbox{$0.3\pm0.3$} in
\mbox{\deltalogmbh}\ at redshift unity. 

CO emission lines afford an opportunity to study QSO host galaxies at higher redshifts.
Shields et al. (2006) used $\sigma_\mathrm{CO} \equiv
\mathrm{FWHM(CO)}/2.35$ together with \mbh\ derived from the broad
\mgii\ and  
\civ\ lines to assess the \mbhsigstar\ relationship in QSOs up to $z =
6.4$.  They find large black holes \mbox{($\sim10^{9.5}~\msun$)} that are at
least an order of magnitude larger than expected for the CO line width
(see Figure \ref{f:dmbh}).  These giant black holes in the early universe 
evidently reside in comparatively modest galaxies, as found for the case of
SDSS~J1148+5251 ($z = 6.4$) by Walter et al. (2005).

Figure 1 summarizes a variety of measurements of \deltalogmbh\ as a
function of redshift.  Included are a number of $z\sim2$ QSOs from
Netzer et al. (2004) and Shemmer et al. (2004), as selected by Bonning
et al. (2005).  The six radio-loud objects have a mean $\deltalogmbh$
of $-0.4$.  Also shown are nine QSOs from the study of Sulentic et
al. (2004).  For these, we have measured the \oiii\ width from the VLT
spectra of Sulentic et al., kindly provided by P. Marziani (personal
communication).   We made a direct measurement of the FWHM of the
$\lambda5007$ line for those objects having adequate \oiii\ intensity
and not having excessive  \feii\ emission, keeping nine of
17 objects.  We 
measured the continuum flux at 5100~\AA\ rest wavelength from the
spectra, derived the luminosity (for our cosmology), and calculated
\mbh\ from Equation \ref{e:mbh} taking FWHM({\hbeta}) from Sulentic et al.
The VLT measurements have a mean $\deltalogmbh = 0.31$, or 0.48
including only the radio-quiet objects.   Altogether, these results
are consistent with the conclusion of Shields et al. (2003) that there
is relatively little evolution of the \mbhsigstar\ relationship
between $z\sim2$ and the present, albeit with considerable scatter. 


\begin{table}[h]
\caption{Width of narrow \oiii\ line in QSOs measured from the spectra
  by Sulentic et al. (2004), along with \mbh\ derived from Equation
  \ref{e:mbh} (see text). Absolute luminosity $\nu L_\nu$ is given at
  rest wavelength $\lambda5100$ for our adopted cosmology.  HE 0003,
  0005, 0454, and 2349 are radio-loud (Sulentic et al. 2004).  } 
\label{VLT}
\begin{minipage}{\columnwidth}
\centering
\begin{tabular}{cccccc}
\hline
\hline

QSO         &       $z$ &      $\log\,\sigthree$ &  $\log\,\nu L_\nu$  &  $\log\,\mbh$ &  $\deltalogmbh$  \\
\hline 

HE 0003   &    1.077  &               2.41  &              46.12  &               9.26  &  0.68   \\
HE 0005   &    1.412  &               2.73  &              46.25  &               9.40   &  -0.45  \\
HE 0048   &    0.847  &               2.41  &              45.49  &               9.23   &  0.64   \\
HE 0248   &    1.536  &               2.76  &              47.49  &               9.73  &  -0.25   \\
HE 0331   &     1.115  &               2.34  &              46.40  &               9.41  &  1.11   \\
HE 0454   &     0.853  &               2.18  &              45.83  &               8.71  &  1.06   \\
HE 2340   &     0.922  &               2.34  &              46.18  &               9.07  &  0.77   \\
HE 2349   &     1.604  &               2.83  &              46.36  &               9.35  &  -0.91 \\
HE 2355   &     2.382  &               2.67  &              46.71  &               9.77  &  0.15  \\
Average   &      1.305  &              2.52   &              46.31  &              9.33   &  0.31  \\  

\hline
\end{tabular}
\end{minipage}
\end{table}

\section{Does \oiii\ track \sigstar\ in AGN?}
\label{o3}

\begin{figure}[!t]
    \includegraphics[width=\columnwidth,angle=0]{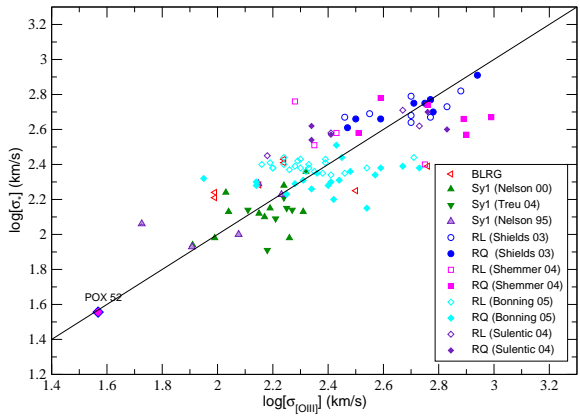}
   \caption{
Correlation between \sigthree\ and \sigstar\ for AGN, from Bonning et al.
(2005) with the addition of POX~52 (Barth et al. 2004) and the VLT
data from Table 1. 
See Bonning et al. for details and references.  This figure follows
Bonning et al. in using a slightly different formula for \mbh\ from
Equation \ref{e:mbh}. 
}
   \label{f:sigsig}
\end{figure}

We noted above some indications that \sigthree\ may be a useful
surrogate for \sigstar\ in a statistical sense.   Figure
\ref{f:sigsig}, based on Bonning et al. (2005), compares \sigthree\
with \sigstar\ in a wide variety of AGN.  Because of the scatter in
\sigthree\ at fixed \sigstar, a wide dynamic range in \sigstar\ is
needed to clarify the overall trend.  For lower luminosity AGN
(Seyfert galaxies), direct measurements of \sigstar\ are used.  For
QSOs, in which \sigstar\ is difficult to measure, \sigstar\ is
inferred from \mbh\ and Equation \ref{e:tremaine}.  We include here
the VLT results of Table 1 and, at very low \sigstar\, the dwarf
Seyfert galaxy POX~52 (Barth et al. 2004).   There is a clear trend of
increasing \sigthree\ with \sigstar, consistent with the idea that
\oiii\ is a valid, if noisy, surrogate.

Greene \& Ho (2005) compare \sigstar\ with the widths of several
narrow emission lines in a sample of narrow-line Seyfert galaxies from
SDSS.  They find that \sigtwo\ agrees in the mean with \sigstar.
However, \sigthree\ exceeds \sigstar\ by $\sim0.13$ dex unless a
correction is made for the extended blue wing of the \oiii\ profile.
In contrast, Salviander et al. (2006) find that \sigthree\ and
\sigtwo\ agree within a few hundredths of a dex in their sample of
SDSS QSOs, without any correction for the blue wing.  It is important
to clarify the effect of the blue wing on \sigthree\ in various
classes of AGN, and how best to correct for it.

\section{Homelessness Amongst the Largest Black Holes}
\label{homeless}

Several authors have noted that the largest black hole masses,
inferred in the most luminous QSOs from broad line widths or the
Eddington limit, exceed the largest values of \mbh\ found in the local
universe, and the implied values of \sigstar\ by
Equation~\ref{e:tremaine} exceed the largest \sigstar\ values in
galaxies (Netzer 2003; Wyithe \& Loeb 2003; Shields \& Gebhardt 2004).
If the black hole masses are correct, this implies a breakdown of the local
\mbhsigstar\ relationship at high \mbh. 

McLure \& Dunlop (2004), in a study of SDSS QSOs, find values of \mbh\
up to $\sim10^{10}~\msun$, but they dismiss the largest values as
resulting from the scatter in deriving \mbh\ from expressions like
Equation~\ref{e:mbh}.  Thus they conclude that black holes in QSOs are
consistent with the maximum mass $\mbh\sim10^{9.5}~\msun$ found in
giant elliptical galaxies, such as M87 (Tremaine et al. 2003).  
There is indeed scatter in the BLR radius as a
function of luminosity (Kaspi et al. 2000), which underlies
Equation~\ref{e:tremaine}.  However, we argue here that the large
values of \mbh\ 
in the most luminous quasars are likely real.  In the QSO sample of
Shields et al. (2003),  there are eight radio-quiet objects with $\log\, \nulnu \geq
46.6$; and of these, five have $\log\,\mbh \geq 9.7$, substantially
exceeding the largest \mbh\ in local galaxies.   Here we are dealing
with a majority of the objects of a class selected by a criterion that
does not involve the scatter in Equation \ref{e:mbh}.  Thus, one
cannot appeal to scatter to dismiss the large values.  These six objects
have an average $\log\,L/\led \approx -0.1$, suggesting that \mbh\
cannot be much less than derived from the \hbeta\ width.    

What is the present-day density of these giant black holes?
From the QSO luminosity function of Boyle et al. (2003),
we estimate the space density of QSOs with
$\log\,\nulnu > 46.6$ to be \mbox{$\sim6~\mathrm{Gpc}^{-3}$} (comoving) at $z = 2$.
Since they are nearly at the Eddington limit, we take their lifetime 
to be the Salpeter e-folding time of 50 million years
(efficiency $0.1c^2$).  Applying the above fraction of 5/8 and taking
 an effective QSO epoch of 3 billion years 
(Warren, Hewitt, \& Osmer 1994), we find the density 
of relic black holes over 5 billion \msun\ to
be $\sim10^{2.3}~\mathrm{Gpc}^{-3}$.  

By the local \mbhsigstar\ and $\mbh-M_\mathrm{bulge}$ relationships
(Kormendy \& Gebhardt 2001), a black hole of 5 billion \msun\ corresponds to 
\mbox{$\sigstar \approx 500~\kmps$} and $M_\mathrm{bulge} \approx 10^{12.6}~\msun$.
The largest  local giant ellipticals (e.g., M87) have
$\sigstar \approx 350~\kmps$ (Faber et al. 1997), and the
velocity dispersion function of SDSS galaxies ends at $\sim400~\kmps$
(Sheth et al. 2003).   Bernardi et al. (2006) find only two or three
candidate galaxies  in SDSS with
$\sigstar > 500~\kmps$  in a volume $\sim0.5~\mathrm{Gpc}^3$ among the objects
for which they find the evidence for superposition to be weakest.  
The nearest black hole with mass 
$\geq 10^{9.7}~\msun$ should be at a distance of $\sim100$~Mpc and redshift of
\mbox{$\sim7000~\kmps$}.  Wyithe (2006) and Wyithe \& Loeb (2003) reach a
similar conclusion based on the assumption that the most luminous
QSOs shine at the Eddington limit.  This distance corresponds to the
largest cD galaxies, such as 
NGC~6166 in Abell~2199 and NGC~7720  in Abell~2634.   Such galaxies
may be a logical place to look for a 5 billion solar mass hole.
However, such a black hole in NGC 6166  or NGC 7720 would violate Equation
\ref{e:tremaine}, since these galaxies have central velocity
dispersions $\sim350~\kmps$ (Tonry 1984).

The space density of galaxy clusters with $\sigma_v~>~500~\kmps$
(Bahcall et al. 2003) exceeds by an order of magnitude our derived
density of black holes over 5 billion solar masses.   Ample dark matter halos
at this velocity dispersion exist in the modern universe, 
but not individual galaxies.   Evidently,  at
this value of \sigstar, the physics of baryon assembly is such that giant black
holes can form in the early universe but the growth of their host
galaxies is stunted.  Perhaps this involves the disruptive effect of the
QSO luminosity (Benson et al. 2003; Wyithe \& Loeb 2003; di Matteo et
al. 2005).  The giant holes in the CO quasars discussed above appear
destined to remain in galaxies of comparatively modest proportions. 

We thank K. Gebhardt, J. Greene, and B. Wills for useful discussions.
EWB acknowledges support from a Chateaubriand Fellowship and a Pierre
and Marie Curie Fellowship.  This work was supported by Texas
Advanced Research Program grant 003658-0177-2001;  by the National
Science Foundation under grant AST-0098594; and by NASA under grant
GO-09498.04-A from the Space Telescope Science Institute.


\begin{thebibliography}{}


\bibitem[Abazajian et al.(2005)]{2005AJ....129.1755A} Abazajian, K., et
al.\ 2005, AJ, 129, 1755

\bibitem[Bahcall et al.(2003)]{Bahcall03} Bahcall, N. A., et al. \ 2003, ApJ, 585, 182

\bibitem[Barth et al.(2004)]{Barth04} Barth, A. J., Ho, L. C., Rutledge, R. E., 
\& Sargent, W. L. W. 2004, ApJ, 607, 90

\bibitem[Benson et al.(2003)]{Benson03} Benson, A. J., et al. \ 2003, ApJ, 599, 38

\bibitem[Bernardi(2006)]{Bernardi06} Bernardi, M. 2006, AJ, in press,
astro-ph/0510696

\bibitem[Bonning et al.(2005)]{2005ApJ...626...89B} Bonning, E.~W., 
et al.\ 2005, ApJ, 626, 89 

\bibitem[Boyle et al.(2000)]{Boyle00} Boyle, B.~J., et al.\ 2000, 
MNRAS, 317, 1014

\bibitem[Combes(2005)]{Combes(2005)} Combes, F. 2005, astro-ph/0505463

\bibitem[Croton(2005)]{Croton05} Croton, D. J. 2005, astro-ph/0512375

\bibitem[Di Matteo et al.(2005)]{2005Natur.433..604D} Di Matteo, T., 
Springel, V., \& Hernquist, L.\ 2005, Nature, 433, 604 

\bibitem[Faber et al. (1997)]{Faber97} Faber, S. M., et al. 1997, AJ, 114, 1771

\bibitem[Ferrarese \& Ford(2004)]{ff04} Ferrarese,
L.~\& Ford, H..\ 2004, Space Science Rev. 116, 523

\bibitem[Ferrarese \& Merritt(2000)]{2000ApJ...539L...9F} Ferrarese,
L.~\& Merritt, D.\ 2000, ApJL, 539, L9

\bibitem[Gebhardt et al.(2000)]{2000ApJ...539L..13G} Gebhardt, K., et
al.\ 2000, ApJL, 539, L13

\bibitem[Greene \& Ho(2005)]{gh05} Greene, J. E., \& Ho, L. C. 2005, ApJ 627,721

\bibitem[Haiman \& Quataert(2004)]{hq04} Haiman, Z., \& Quataert, E. 2004,
in Supermassive Black Holes in the Distant Universe, ed. A. J. Barger,
Astrophysics and Space Science Library (Dordrecht: Kluwer), Vol. 308, p. 147

\bibitem[Kaspi et al.(2000)]{k00} Kaspi, S., et al.\
2000, ApJ, 533, 631

\bibitem[Kaspi (2005)]{k05} Kaspi, S., et al. 2005. ApJ, 629, 61

\bibitem[Kormendy \& Gebhardt(2001)]{kg01} Kormendy,
J.~\& Gebhardt, K.\ 2001, AIP Conf.~Proc.~586: 20th Texas Symposium on
Relativistic Astrophysics, 586, 363

\bibitem[McLure \& Dunlop(2004)]{2004MNRAS.352.1390M} McLure, R.~J., \& 
Dunlop, J.~S.\ 2004, MNRAS, 352, 1390 

\bibitem[Nelson \& Whittle(2000)]{nelson96} Nelson, C.~H., \& Whittle, M. 1996, ApJS,
99, 67

\bibitem[Nelson(1996)]{nelson00} Nelson, C.~H.\ 2000, ApJ
544, L91

\bibitem[Netzer(2003)]{netzer03} Netzer, H.\ 2003, ApJL, 583, 
L5

\bibitem[Netzer et al.(2004)]{netzer04} Netzer, H., et al. \ 2004, ApJ, 614, 558

\bibitem[Salviander(2006)]{ss06} Salviander, S., Shields, G.~A.,  Gebhardt, K.,
\& Bonning, E.~W. 2006, submitted.

\bibitem[Shemmer et al.(2004)]{shemmer04} Shemmer, O., et al. \ 2004, ApJ, 614, 547

\bibitem[Sheth et al.(2003)]{sheth03} Sheth, R. K. \ 2003, ApJ, 594, 225

\bibitem[Shields \& Gebhardt (2004)]{shields04} Shields, G.~A., \&
Gebhardt, K., 2004, Bull. AAS, 204.6002S

\bibitem[Shields et al.(2003)]{s03} Shields, G.~A.,
et al.\ 2003, ApJ, 583, 124

\bibitem[Shields(2006)]{shields06} Shields, G.~A., Menezes, K. L.,
Massart, C. A., \& Vanden Bout, P. 2006, ApJ, in press, astro-ph/0512418

\bibitem[Sulentic et al.(2004)]{Sulentic04} Sulentic, J. W., et al. 2004, AstrAp, 423, 121

\bibitem[Tonry(1984)]{tonry02} Tonry, J. L.
1984, ApJ, 279, 13

\bibitem[Tremaine et al.(2002)]{Tremaine02} Tremaine, S., et al.\ 
2002, ApJ, 574, 740

\bibitem[Walter et al. (2004)]{Walter04} Walter, F., et al. 2004, ApJL, 615, L17

\bibitem[Warren et al.(1994)]{Warren94} Warren, S. J., Hewitt, P. C., \& Osmer, P. S. 1994,
ApJ, 421, 412

\bibitem[Wyithe(2006)]{Wyithe06} Wyithe, J. S. B. 2006, MNRAS, 365, 1082

\bibitem[Wyithe\&Loeb(2003)]{Wyithe03} Wyithe, J. S. B., \& Loeb, A. 2003, ApJ, 595, 614

\end{thebibliography}
\end{document}